\documentclass[figures,cite]{epl}
\usepackage{epsfig}

\title{Proximity effects in the superconductor / heavy fermion bilayer system Nb / CeCu$_6$.}
 \shorttitle{Proximity effects in Nb / CeCu$_6$ }
\author{A. Otop\inst{1,2} \and R. W. A.  Hendrikx\inst{1} \and M. B. S.
Hesselberth\inst{1} \and C. Ciuhu\inst{3} \and \\ A. Lodder\inst{3} \and J.
Aarts\inst{1}}
 \shortauthor{A. Otop \etal}
 \institute{
 \inst{1} Kamerlingh Onnes Laboratory, Leiden University, P.O. Box 9504, 2300 RA Leiden,
 the Netherlands\\
 \inst{2} Institut f\"ur Metallphysik und Nukleare Festk\"orperphysik, Technische
Universit\"at Braunschweig, Mendelssohnstr. 3, 38106 Braunschweig,
Germany. \\
  \inst{3} Faculty of Sciences, Vrije Universiteit, De Boelelaan 1081, NL-1081~HV~Amsterdam, ~The
Netherlands\\ }

 \pacs{71.27.+a,}{}
 \pacs{74.50.+r}{}

\begin{document}

\maketitle \today \\

\begin{abstract}
We have investigated the proximity effect between a superconductor (Nb) and a
'Heavy Fermion' system (CeCu$_6$) by measuring critical temperatures $T_c$ and
parallel critical fields $H_{c2}^{\parallel}$(T) of Nb films with varying
thickness deposited on 75~nm thick films of CeCu$_6$, and comparing the results
with the behavior of similar films deposited on the normal metal Cu. For Nb on
CeCu$_6$ we find a strong decrease of $T_c$ with decreasing Nb thickness and a
finite critical thickness of the order of 10 nm. Also, dimensional crossovers
in $H_{c2}^{\parallel}$(T) are completely absent, in strong contrast with
Nb/Cu. Analysis of the data by a proximity effect model based on the
Takahashi-Tachiki theory shows that the data can be explained by taking into
account both the high effective mass (or low electronic diffusion constant),
{\it and} the large density of states at the Fermi energy which characterize
the Heavy Fermion metal.
\end{abstract}

\section{Introduction}
When a superconductor (S) is brought into contact with a non-superconducting
conductor (N), superconductivity leaks into that material by the proximity
effect \cite{degennes64}. For the thermodynamic properties of the S/N bilayer
such as the critical temperature $T_c$ or the upper critical field $H_{c2}$,
this can be modelled as a spatial variation of the superconducting pair
density. The spatial variation mainly depends on the electron diffusion
constants for the S- and the N-metal, on the transparency of the interface for
electrons and Cooper pairs, and on the pair breaking mechanisms on the N-side
of the interface. Until now, two classes of N-materials have been investigated.
The first is formed by simple metals such as Cu or Au, where the physics is
mostly understood. Pairs are broken at finite temperatures by thermal
fluctuations, leading to dephasing of the constituents of the induced Cooper
pair. This is translated into a temperature-dependent characteristic length
over which superconducting correlations penetrate, the 'normal metal coherence
length' $\xi_N$. Since $\xi_N = \sqrt{(\hbar D_N) /(2 \pi k_B T)}$ ($D_N$ is
the diffusion constant of the N-metal, $T$ is the temperature, the other
symbols have their usual meaning), this length can become large at low
temperatures. In the absence of other pair breakers, $T_c$ of a bilayer with
finite N-layer thickness $d_N$ will be finite. The other class of materials is
formed by ferromagnets (F) such as Fe; pair breaking is due to the exchange
interaction $E_{ex}$ which acts on the spins of the Cooper pair. For strong
magnets ($E_{ex} >> k_B T_c$) it results in a temperature-independent $\xi_F =
\sqrt{(\hbar D_F) / (2 \pi E_{ex})}$ with a characteristic value of only a few
nanometer. In this case, superconductivity can already be quenched at a finite
value for $d_S$, called the critical thickness $d_S^{cr}$
\cite{muehge96,aarts97}.  \\
Little attention has yet been paid to N-layers where the electronic ground
state is dominated by many body correlations, such as in a Heavy Fermion (HF)
metal. Basically, HF metals consist of a lattice of atoms with localised
(f-)electrons, where the magnetism is quenched by a coherent Kondo effect. This
leads to a strong peak in the DOS near the Fermi energy with small energy width
(for relevant reviews, see \cite{hewson93,nieuwen95}). The consequences for the
low temperature physics can be phenomenologically described in terms of Landau
Fermi liquid theory by a mass renormalization of the charge carriers. The
ensuing large effective mass $m^*$ can be directly observed in e.g. the
specific heat of the system, $c_v$, where the linear term $\gamma = c_v/T
\propto m^*/m_e$ (with $m_e$ the bare electronic mass) can be up to two orders
of magnitude larger than in normal metals. Since the magnetic moments are often
not completely quenched, HF systems can also show magnetic order at low
temperatures. For the proximity effect in an S/HF system, several scenarios are
possible. In the spirit of the Fermi liquid approach, the HF metal can be
considered a normal metal with a large mass, or equivalently a low Fermi
velocity and therefore a small diffusion constant. Also, additional pair
breaking mechanisms may be present due to the strong electron-electron
interactions or the residual moments. On the other hand, the interface
transparency may be decreased due to the mismatch in Fermi velocities, which
would shield the superconductor from the HF metal and counter the other two effects. \\
In order to investigate these questions, we have performed a comparative
proximity effect study of the thin film systems Nb/CeCu$_6$ and Nb/Cu. CeCu$_6$
is a well-known HF system, with an extremely large value of $\gamma \approx$
1.6~J/(mole K$^2$) \cite{schlager93} and no magnetic order down to the
mK-regime, making it a good model system for this study. Nb/Cu is a much
studied proximity system which shows one feature of particular interest, namely
a Dimensional Crossover (DCO) from three-dimensional (3D) to two-dimensional
(2D) behavior in the temperature dependence of the critical field parallel to
the layers $H_{c2}^{\parallel}$ \cite{banerjee84,chun84}. The DCO occurs when
the N-layer thickness becomes of the order of $\xi_N$, and can be modelled
quite accurately. In particular, recent calculations by Ciuhu and Lodder based
on the Takahashi-Tachiki theory for metallic multilayers \cite{taka86}, and
including a finite interface resistance $R_B$, showed that quantitative
agreement between theory and experiment can be found in the case of Nb/Cu for
very reasonable values of the different parameters \cite{ciuhu01}. Here we
present data on $T_c(d_S)$ and $H_{c2}^{\parallel}(T)$ for bilayers Nb/Cu and
Nb/CeCu$_6$ and compare them with similar calculations. We show that the HF
system can be treated as a normal metal, with {\it both} the low diffusion
constant (Fermi velocity) {\it and} the high DOS as necessary ingredients to
explain the observed behavior.

\section{experimental}
 Sets of CeCu$_{6}$ films and bilayers of {\it sub}/CeCu$_{6}$/Nb ({\it sub} denotes
the substrate) were fabricated by DC-magnetron sputtering in an ultra high
vacuum (UHV) system with a background pressure of the order of $5 \times
10^{-10}$ mbar, and an Ar sputtering pressure of $2.5 \times 10^{-3}$ mbar.
Crystalline CeCu$_{6}$ was grown as reported before \cite{groten01}, at a
temperature of 350~$^{\circ}$C, using Si-substrates with amorphous
Si$_{3}$N$_{4}$ buffer layers to prevent Cu diffusion at those temperatures.
\begin{figure}
\epsfig{file=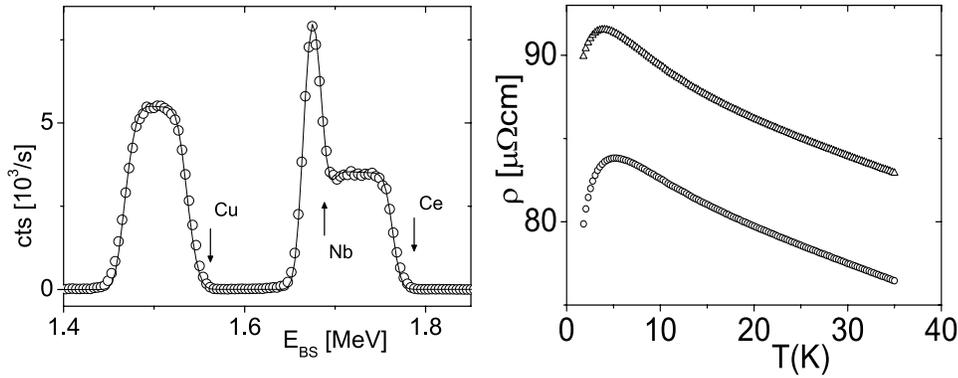,width=13cm}
 \caption{Materials characteristics of the CeCu$_6$ films. (a) RBS
spectrum (nr. of counts versus backscatter energy E$_{BS}$) taken with
$^4$He-ions of 2~MeV on a sample Si/Si$_{3}$N$_{4}$/(75 nm CeCu$_{6})$/(15 nm
Nb). The different elements are indicated. The thin smooth line is a fit to the
measured curve. (b) Specific resistance as function of temperature for single
CeCu$_6$ films of 75~nm (o) and 100~nm ($\triangle$) .}\label{fig1-rbsrho}
\end{figure}
The Nb was deposited on top of the CeCu$_6$ after cooling the substrate holder
with cold nitrogen gas to close to room temperature. Composition, thickness and
crystallinity of the films were determined by Rutherford BackScattering (RBS)
measurements together with X-ray diffraction measurements  at low and high
angles. The RBS measurements show good agreement with the expected
stoichiometry for CeCu$_{6}$ and no diffusion is found either of Ce or Cu into
the substrate or of Nb into the CeCu$_{6}$. Fig.~\ref{fig1-rbsrho}{\it a} shows
part of the RBS spectrum for CeCu$_{6}(75~nm)$ / Nb(15~nm) on a
Si/Si$_{3}$N$_{4}$ substrate and a fit of the data without taking any diffusion
into account. Bilayers and trilayers of {\it sub}/Cu/Nb and Cu/Nb/Cu were grown
in a different UHV system with similar background pressure and sputtering conditions.
In order to compare results, $d_{Nb}$ in the trilayers was taken two times $d_{Nb}$
in the bilayers, which yields equal conditions for the superconducting order
parameter in the middle of the film (for the trilayer) and at the vacuum interface
(for the bilayer).\\
Resistance measurements were performed in standard 4-point geometry on
lithographically patterned samples with bridge widths of 200~$\mu$m and a
distance of 1.2~mm between the voltage contacts. The electrical resistivity
$\rho_{CeCu6}$ as function of temperature for two single films is shown in
Fig.~\ref{fig1-rbsrho}{\it b} and behaves as reported before \cite{groten01},
with a clear maximum in $R$(T) at T$_{max} \approx$ 5~K, similar to what is
found for bulk material \cite{amato85}, and a residual resistivity of the order
of 80~$\mu \Omega cm$. We decided to use 75 nm thick CeCu$_{6}$ layers, which
with T$_{max}$=4 K suggest only little deviation from the bulk properties.

\section{Results and discussion}
\begin{figure}
\epsfig{file=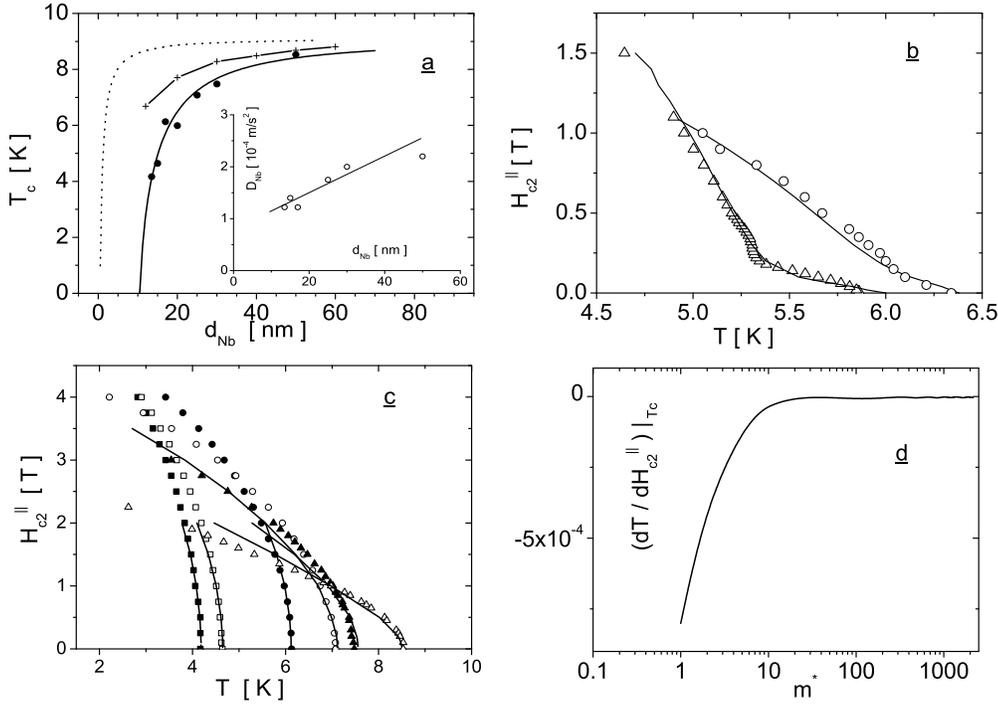,width=14.5cm} \caption{(a) $T_c$ as function of Nb
thickness $d_{Nb}$ for bilayers {\it sub}/CeCu$_{6}$~/Nb ($\bullet$)and single
Nb films (+). The drawn line is a fit using the proximity effect model. The
dotted line is a model calculation with $N_{CeCu6}$ = $N_{Cu}$. The inset shows
the behavior of $D_{Nb}$ as found from the fit. (b) Parallel critical field
$H_{c2}^{\parallel}$ as function of $T$ for a bilayer {\it sub}/Cu/Nb
($\triangle ; d_{Nb}$ = 15 nm) and a trilayer Cu/Nb/Cu (o; $d_{Nb}$ = 30 nm).
Drawn lines are model fits. (c) $H_{c2}^{\parallel}$ as function of $T$ for
bilayers {\it sub}/CeCu$_6$/Nb with thicknesses $d_{Nb}$ = 13.5 nm (lowest
$T_c$), 15 nm, 17 nm, 25 nm, 30 nm, 50 nm. Drawn lines are model fits. (d)
Model calculation of the inverse slope of $H_{c2}^{\parallel}$ at $T_c$ versus
the effective electron mass $m^*$.}\label{fig2}
\end{figure}

The dependence of $T_c$ on $d_{Nb}$ for the bilayer set {\it sub}/CeCu$_{6}$(75
nm)/Nb($d_{Nb}$) is shown in Fig.~\ref{fig2}{\it a} and compared to that of
single Nb films, prepared in both UHV systems. For the Nb films a slight
suppression is witnessed, usually found in the case of Nb, and caused by a
mixture of different effects such as oxidation through the unprotected top
layer, a small proximity effect with the substrate, and lifetime broadening due
to growth disorder \cite{park85}. For the S/HF bilayers, the suppression is
much stronger, with a critical thickness $d_{cr}$ for onset of
superconductivity reached around 12~nm. This would be equivalent to 24~nm in a
trilayer configuration, and is of a similar magnitude as found in
superconductor/ferromagnet systems such as V/Fe and Nb/Fe
\cite{aarts97,geers01}. The result shows immediately that the interface does
allow particle exchange, so the supposed huge Fermi velocity mismatch in the
system of order 10$^3$ does not lead to a highly reflective interface. It also
shows that the coherence length in the HF-material $\xi_{HF}$ must be small,
leading to the strong suppression. \\
Another indication of behavior deviating from simple S/N systems comes from the
parallel critical field H$_{c2}^{\parallel}(T)$. To demonstrate the difference,
Fig.~\ref{fig2}{\it b} shows H$_{c2}^{\parallel}(T)$ for a bilayer of {\it
sub}/Cu(75~nm)~/ Nb(15 nm) and a trilayer of Cu(75~nm)~/ Nb(30~nm)~/ Cu(75~nm),
to be compared to the data of the S/HF bilayers given in Fig.~\ref{fig2}{\it
c}.
Decreasing $T$ near $T_c$, H$_{c2}^{||}(T)$ is linear for the Nb/Cu samples,
followed by a kink and $\sqrt{1-T/T_c}$~-like 2D behavior. The kink signals the
well-known DCO \cite{banerjee84,chun84}, usually observed in multilayers, but
also present in tri- and bilayers. In strong contrast, the S/HF bilayers do not
show a DCO but only 2D-behavior for all $d_{Nb}$. Qualitatively, this again
indicates a small value for $\xi_{HF}$ : the superconducting order parameter
does not penetrate sufficiently far into the HF metal to yield a coupled
system. \\
To make more quantitative statements, we analyzed the data by means of model
calculations based on the Takahashi-Tachiki formalism. Details are given in
ref.~\cite{ciuhu01}, but we reiterate the main elements in order to introduce
the different parameters. The formalism solves the equation for the pair
potential $\Delta({\bf r})$ with a space-dependent coupling constant $V({\bf
r})$ for small $\Delta$ close to $T_c$ :
\begin{equation}
\Delta({\bf r}) =  V({\bf r}) k_B T \sum_{\omega} \int d^3{\bf r'}
Q_\omega({\bf r,r'}) \Delta({\bf r'})
\end{equation}
in which $Q_\omega$ is an integration kernel still to be determined, $V({\bf
r})$ is the BCS coupling constant and the other symbols have their usual
meaning; the summation runs over the Matsubara frequencies. In the dirty limit,
$Q_\omega$ can be shown to obey :
\begin{equation}
[\; 2|\omega| - \hbar D({\bf r}) (\; {\bf \nabla} - \frac{2ie}{\hbar c} {\bf
A(r)} \;)^2 \;]\; Q_\omega({\bf r,r'})  \;=\; 2 \pi N({\bf r}) \delta({\bf
r}-{\bf r'})
\end{equation}
with $N({\bf r})$ the DOS at the Fermi energy and $D({\bf r})$ the diffusion
constant. These equations are complemented with boundary conditions for
$\Delta$ at the interface which parametrize a possible barrier encountered by
the electrons through a boundary resistivity $R_B$. In the calculations we use
fixed values for the density of states ratios $N_{Cu} / N_{Nb}$ = 0.16 and
$N_{CeCu6} / N_{Nb}$ = 320. The former value was also used in
ref.~\cite{ciuhu01} and derives from ref.~\cite{banerjee82}; the latter value
is constructed from the former by the ratio $\gamma_{CeCu6} / \gamma_{Cu}$ =
2000. The value of $T_{c,Nb}$ was fixed at 9.2~K, except for the Nb/Cu bilayer
with $d_{Nb}$ = 15~nm where it was 8.2~K. This reflects the fact that $T_c$ for
thin Nb-layers starts to decrease, as explained above. Fitted were the
different diffusion constants $D_N$ and boundary resistances $R_B$. The results
of the fits for the critical field data are shown in Fig.~\ref{fig2} as solid
lines. The parameter values are given in Table~\ref{table-values}.
\begin{table}
\centering
\begin{tabular}{|c|c|c|c|c|} \hline
  metal & $d_N$ &  $D_{N}$  & $D_{Nb}$  &  $R_B$   \\
  &       &    &          &          \\ \hline
Cu (bi) & 15  &  240  &  1.4  &  177   \\
Cu (tri) & 30  &  220  &  2.9  &  252   \\
CeCu$_6$ & 13.5  &  0.15  &  1.2  &  324   \\
         & 15    &  0.19  &  1.4  &  338  \\
         & 17  &  0.11    &  1.2  & 350 \\
         &  25  &  0.11  &  1.75  & 252 \\
         &  30  &  0.11  &  2.  & 252   \\
         &  50  &  0.11  &  2.2  & 324  \\
         \hline
\end{tabular}
\caption{Values for the normal metal thickness $d_N$ (in nm), the fitted values
for the diffusion constants $D_{Cu}$, $D_{CeCu6}$, $D_{Nb}$ (in $10^{-4} m^2 /
s$) and the interface resistance $R_B$ (in $10^{-8} \mu \Omega cm^2$ ) for the
different samples discussed in the text} \label{table-values}
\end{table}
For the two Nb/Cu samples, the fitted values are in very reasonable agreement
with the numbers found by fitting the data of Chun {\it et al}
\cite{chun84,ciuhu01}. The values for $D_{Nb}$ are roughly equal, the values
for $D_{Cu}$ are slightly higher, which probably reflects the difference in
preparation conditions, and also the values for $R_B$ are very similar
\cite{note1}. For the Nb/CeCu$_6$ samples it can be noted that $D_{Nb}$
increases slowly and more or less linearly with increasing $d_{Nb}$; using this
linear variation, which is plotted in the inset of Fig.~\ref{fig2}{\it a}, and
values of $D_{CeCu6}$ = 0.1 $\times 10^{-4} m^2 s^{-1}$, $R_B$ = 324 $\times
10^{-8} \mu \Omega cm^2$ we calculated the behavior of $T_c(d_S)$ as a
consistency check. The agreement, shown in Fig.~\ref{fig2}{\it a}, is equally
satisfactory. The most interesting values from the fits are of course those for
$D_{CeCu6}$, which are much lower than those for Cu. Assuming a Fermi velocity
of the order of $10^3$ m/s, a value for $D_{CeCu6}$ of $0.1 \times 10^{-4}$
m$^2$/s yields a mean free path $l_e$ of 3~nm, which is not surprising in view
of the strongly granular nature of the films. Still, a low value for $D_N$ by
itself does not necessarily reflect the heavy Fermion character : if CeCu$_6$
is taken as a Cu matrix strongly diluted by a small amount of Ce atoms, with a
mean free path of the order of the interatomic distance, $D$ would also be very
small, of order $10^{-4}$~m/s. However, as we show now, just a low value for
$D_N$ cannot describe the measurements. To demonstrate this, we calculated
$T_c(d_S)$ for Nb/CeCu$_6$ with the parameters given above, but with the
DOS-ratio for Nb/Cu, thereby mimicking a very dirty but otherwise standard
N-metal. The result, shown in Fig.~\ref{fig2}{\it a} as dotted line, shows that
$T_c$ now drops at much lower $d_{Nb}$. This can be understood by realizing
that the low diffusion constant inhibits penetration of Cooper pairs in the
N-metal and therefore also inhibits pair breaking, leading to a smaller amount
of suppression of $T_c$. What makes the difference is the high DOS-value for
CeCu$_6$ : the large number of available states works as a sink for Cooper
pairs which counteracts the low diffusion constant. Our major conclusion is
therefore that both the low $D_N$ {\it and} the high $N_N$ are necessary
ingredients in the description of the data. This leads to the question what the
effective mass needs to be in order to suppress the DCO which is so
characteristic for an S/N system. For this we calculated values for $dT /
dH_{c2}^{\parallel}|_{Tc}$, the inverse parallel critical field slope at $T_c$,
as function of the effective mass $m^* / m_e$ ($m_e$ being the bare electron
mass) as used in the free-electron expressions for $N_N = \frac{m^*
k_F}{\hbar^2 \pi^2}$ and $D_N = \frac{k_F l_e}{3m^*}$, with $k_F$ the Fermi
wave vector. As shown in Fig.~\ref{fig2}{\it d}, low values for $m^*$ yield a
finite value for the inverse slope, signifying 3D behavior and therefore a DCO,
which goes to 0 around $m^*$ = 10, meaning that the DCO has disappeared and 2D
behavior has set in. Clearly, CeCu$_6$ is far
into this regime.\\

Finally, we come back to the difference in proximity effects between F-, and
HF-metals. Using the expression for $\xi_N$ given in the Introduction at a
typical value of $T$ = 5~K and with the fitted values for $D_{CeCu6}$, we find
$\xi_{CeCu6} \approx$ 1.5~nm. This is very similar to values found for strong
ferromagnets, but the physics of the strong suppression of superconductivity
which is found both in the F- and HF-systems is different. In the F-case, the
small value of $\xi_F$ derives from the large pair breaker ($E_{ex}$), in the
HF-case from the small $D_N$. As shown in Fig.~\ref{fig2}a, a low value for
$D_N$ does not yield strong suppression of superconductivity, that is actually
due to the high value for $N_N$. This emphasizes once more that the basic
proximity-effect parameter is not $\xi_{F,HF}$ but rather $\gamma = (\rho_S
\xi_S) / (\rho_{X} \xi_{X})$ with X = F, HF \cite{aarts97,gol94}. Since $\rho_X
\propto (N_X D_X)^{-1}$ and $\xi_X \propto \sqrt{D_X}$ it can easily be seen
that for the F-system $\gamma$ is large due to $E_{ex}$, and for the HF-system
$\gamma$ is large due to $N_{CeCu6}$. For the HF-system, an extra pair breaker
is not needed in the description.

\acknowledgments This work is part of the research program of the 'Stichting
voor Fundamenteel Onderzoek der Materie (FOM)', which is financially supported
by NWO. We would like to thank R. Boogaard for preparing and measuring the
Cu/Nb bilayers. A.O. acknowledges support from the ESF program 'FERLIN'.


\begin{thebibliography}{99}
\bibitem{note1} The values for $R_B$ given in ref.~\cite{ciuhu01} are quoted in
units $\mu \Omega cm$ and were obtained by somewhat arbitrarily multiplying the
fitted value for $R_B$ by a sheet thickness of 10~nm. A value such as 3.17~$\mu
\Omega cm$ in ref.~\cite{ciuhu01} therefore corresponds to 317 $ 10^{-8} \mu
\Omega cm^2$ in the present paper.
\end{thebibliography}
\end{document}